%% file: main.tex
\documentclass[runningheads]{llncs}
\usepackage{comment}
\usepackage{graphicx}
\usepackage{xcolor}
\usepackage{amsmath}
\usepackage{amsfonts}
\usepackage{booktabs}
\usepackage{multirow}
\usepackage{tabularx}
\usepackage{epsfig}
\usepackage{bm}
\usepackage{tablefootnote}
\usepackage{color}

\begin{document}
\title{Integrative Graph-Transformer Framework for Histopathology Whole Slide Image Representation and Classification}
%
%

\author{Zhan Shi$^1$, Jingwei Zhang$^1$, Jun Kong$^2$, and Fusheng Wang$^1$}
\institute{$^1$Stony Brook University, USA\\
$^2$Georgia State University, USA\\
{ \email{\{shzhan, jingwezhang, fushwang\}@cs.stonybrook.edu}\\ \email{jkong@gsu.edu}}}


%

\maketitle              
\begin{abstract}
In digital pathology, the multiple instance learning (MIL) strategy is widely used in the weakly supervised histopathology whole slide image (WSI) classification task where giga-pixel WSIs are only labeled at the slide level. However, existing attention-based MIL approaches often overlook contextual information and intrinsic spatial relationships between neighboring tissue tiles, while graph-based MIL frameworks have limited power to recognize the long-range dependencies.
In this paper, we introduce the integrative graph-transformer framework that simultaneously captures the context-aware relational features and global WSI representations through a novel Graph Transformer Integration (GTI) block. Specifically, each GTI block consists of a Graph Convolutional Network (GCN) layer modeling neighboring relations at the local instance level and an efficient global attention model capturing comprehensive global information from extensive feature embeddings. 
Extensive experiments on three publicly available WSI datasets: TCGA-NSCLC, TCGA-RCC and BRIGHT, demonstrate the superiority of our approach over current state-of-the-art MIL methods, achieving an improvement of 1.0\% to 2.6\% in accuracy and 0.7\%-1.6\% in AUROC.



\end{abstract}

%
\input{chapter/introduction}

\input{chapter/method}
\input{chapter/experiment}
\input{chapter/conculsion}
%
\pagebreak 
\bibliographystyle{splncs04}
\bibliography{ref}

\end{document}

%% file: chapter/introduction.tex
\section{Introduction}
With the significant advance in high-throughput whole slide tissue scanning technology, digital pathology leverages high-quality whole slide images (WSIs) and is an actively developing component in pathology study~\cite{niazi2019digital}. As WSIs are often
giga-pixels by scale and lack of pixel-level annotations, an efficient and effective way to analyze such high-resolution WSIs becomes critical to facilitate cancer diagnosis and prognosis. Due to the remarkable performance, deep-learning based multiple instance learning (MIL) is often employed in such weakly-supervised scenarios where only slide-level labels are available~\cite{hou2016patch,khened2021generalized,kim2019deep,yao2020whole}. By the MIL scheme, each image patch or instance is first encoded as a feature embedding using a pretrained feature extractor~\cite{kim2019deep}. These embeddings are next passed to an aggregator module that compiles embeddings into a comprehensive bag-level representation before the classification~\cite{wang2018revisiting}.

Multiple digital pathology studies in the MIL framework adopt attention mechanisms and achieve promising results with the global WSI representations~\cite{attentionpooling,clam_method}. However, these methods assume that all instances are independent and  thus ignores the critical correlations across different tissue regions. The self-attention mechanism from vision transformers (ViT)~\cite{dosovitskiy2020image} has been used to address this problem, where pairwise similarity scores across all instances are computed~\cite{transmil,chen2022scaling,xiong2023diagnose}. 
However, such pairwise calculation exhibits quadratic complexity, often too demanding to support a large number of input instances for the WSI classification. The Nystr\"{o}m-attention~\cite{xiong2021nystromformer} has been applied to alleviate this problem in TransMIL~\cite{transmil}. It utilizes a subset of landmarks to approximate the self-attention process. Similarly, FlashAttention~\cite{flashattention} achieves the full self-attention ability and uses the IO-aware mechanism to enhance the attention efficiency. 

Besides the correlation across instances, the tissue spatial relationship is crucial for the WSI analysis. However, it is often overlooked in existing MIL studies.
The use of position encoding in Transformers for fixed-length sequences
~\cite{dosovitskiy2020image} can preserve positional information, but it cannot be directly used in the WSI analysis due to the variable lengths of input instance embeddings. To address this, TransMIL~\cite{transmil} employ Convolution Neural Network(CNN) to characterize the spatial information. However, it reorganizes the tissue patches and thus does not accurately reflect the genuine spatial relationships among patches.
Consequently, the inherent potential for spatial arrangement within WSI has not been thoroughly explored.

By contrast, the graph structure is widely known for its intrinsic merit for spatial relationship representations and graph-based MIL methods have increasingly gained attention for the histopathology WSI analysis. 
The Graph Convolution Network (GCN) utilizes a foundational local message-passing mechanism to capture spatial interactions and integrate neighboring instances and provides a cutting-edge graph-based paradigm for the digital pathology study. 
However, such GCN-based frameworks may suffer from over-smoothing~\cite{kreuzer2021rethinking} due to the repeated aggregation of local information, and over-squashing~\cite{alon2020bottleneck} as a result of the increased model depth. 
Moreover, graph-based MIL frameworks exhibit limitations in recognizing long-range dependency.

Recent research has demonstrated that integrating self-attention mechanisms into graph-based approach can effectively mitigate the limitations of message-passing mechanism, such as over-squashing and over-smoothing, thereby enhancing the model's capability for representation~\cite{rampavsek2022recipe}. Furthermore, the application of graph transformers has extended to multi-modal, multi-task, and multi-scale analysis of WSIs~\cite{nakhli2023sparse,zhao2023mulgt,ding2023multi}.
The GTP~\cite{GTP} has been developed for WSI classification which employs a clustering-based mincutpool~\cite{bianchi2020spectral} to bridge GCN and transformer layers. However, the GCN layers in GTP are still prone to over-squashing, and the inevitable information loss from the pooling layer constrains the transformer's capabilities. 

To alleviate these limitations, we develop a novel Integrative Graph-Transformer (IGT) framework for WSI representation and classification.
The core architecture of the IGT framework consists of a sequence of graph transformer integration blocks, where each block integrates a GCN layer for encoding spatial relationships among adjacent instances and a global attention module capturing global WSI representations.  Our framework is able to simultaneously models spatial relationships at the local instance level and long-range pairwise correlations across all instances. 
We demonstrate the efficacy of our method on three public WSI datasets, TCGA-NSCLC, TCGA-RCC and BRIGHT. With extensive testing on these datasets, our IGT framework presents a superior performance to the state-of-the-art methods, achieving a 1.0\% to 2.6\% improvement in accuracy and a 0.7\% and 1.6\% increase in AUROC.

%% file: chapter/method.tex
\section{Method}
\begin{figure*}[t!]
    \centering
    \includegraphics[width=\textwidth]{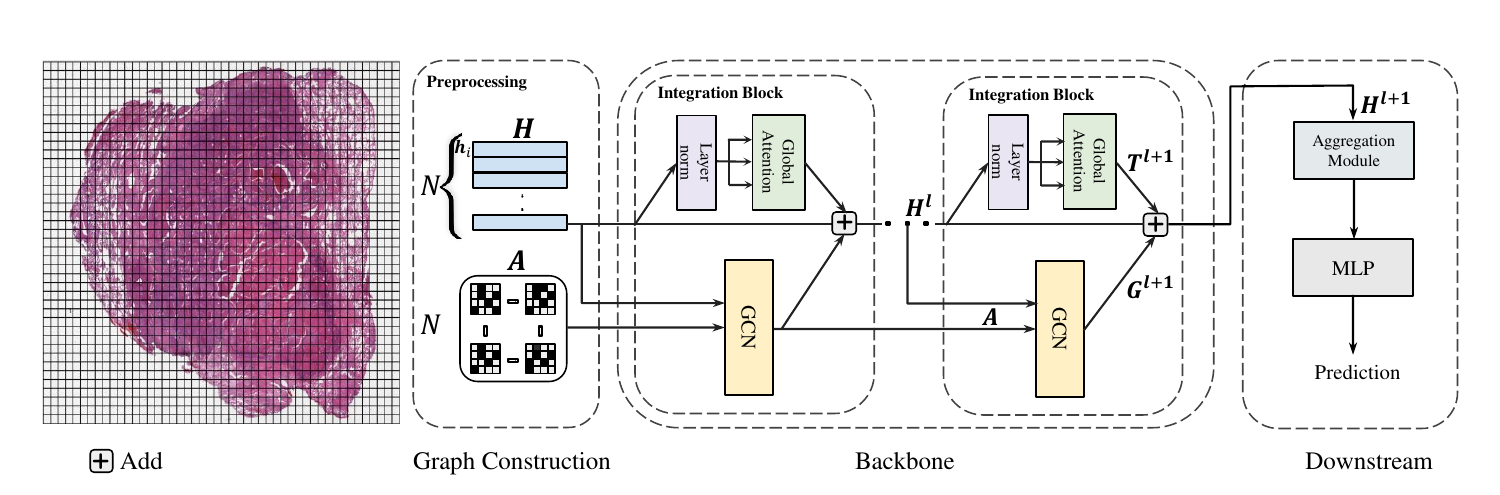}
    \caption{Overview of the proposed method:  The process begins with the graph construction module where a graph representation $G = (\mathbf{H}, \mathbf{A})$ is generated for the subsequent backbone network. Here, the $\mathbf{H}$ is the feature matrix and $\mathbf{A}$ denotes the associated adjacent matrix. Within each $l$-th integration block of the backbone, a global attention layer processes $\mathbf{H}^l$ to produce the feature matrix $\mathbf{T}^{l+1}$, and a GCN layer processes both $\mathbf{H}^l$ and $\mathbf{A}$ to update the graph representation $\mathbf{G}^{l+1}$. Finally, the integrated feature $\mathbf{H}^{l+1}$ from the last block is utilized for prediction in the downstream module.}
\label{fig:method}
\end{figure*}

Illustrated in Fig.~\ref{fig:method}, the proposed IGT framework consists of three key components: graph construction, the backbone, and the downstream process. During the graph construction, feature vectors are extracted, and the corresponding adjacency matrix is created. The backbone module processes this undirected WSI graph representation, serving as an efficient encoder. Finally, the refined features from the last GTI block are provided to the downstream model for classification.

\subsection{Graph Construction}
For each WSI graph $G$ construction, we first partition a WSI into non-overlapping $256\times 256$ tissue region patches/instances. Note that the number of extracted instances $N$ varies for different WSIs. A ResNet50~\cite{he2016deep} model pre-trained on ImageNet is used to encode each instance into a 1024-dimensional feature vector $\{\mathbf{h}_{i}\in\mathbb{R}^{1024}, i=1,2...N\}$. Each feature is regarded as a node in the WSI graph and we assemble instance feature vectors as the node feature matrix $\{\mathbf{H}\in\mathbb{R}^{N\times1024}\}$ for each WSI. To depict the node connectivity in the WSI graph, we preserve the patch spatial coordinates in the WSI and find adjacent nodes by the K-Nearest Neighbor algorithm (i.e. \textit{k-NN}, k=8)~\cite{cover1967nearest}. Thus, we build the WSI graph $G = (\mathbf{H}, \mathbf{A})$ in the Euclidean space, where $\{\mathbf{A}=[A_{i}{}_{j}] ,\mathbf{A}\in \mathbb{R}^{N\times N}\}$ is the adjacency matrix. Its entry $A_{i}{}_{j} =1$ when  there exists a connection between node $i$ and $j$ by the \textit{k-NN} algorithm on node feature representations $(\mathbf{h}_{i}, \mathbf{h}_{j})$. Otherwise, $A_{i}{}_{j}=0$. This graph models the local neighborhood information across the entire WSI.
\subsection{Graph-Transformer Integration Block}


The spatial relationships across tissue instances are crucial for the WSI representation and classification~\cite{ahmedt2022survey}. Therefore, we design the GTI block to concurrently aggregate the local instance relationships and capture long-range pairwise correlations across the entire tissue domain. 


As depicted in the integration block (Fig.~\ref{fig:method}), the $l$-th GTI block operates on the $GCN$ and $GlobalAttention$ layers in parallel and integrates their outputs through a simple summation as follows: 
\begin{align}
\mathbf{G}^{l+1} \ &=\ GCN\left(\mathbf{H}^{l} ,\ \mathbf{A}\right) \label{eq:1}\\
\mathbf{T}^{l+1} \ &=\ GlobalAttn\left(\mathbf{H}^{l}\right) \label{eq:2} \\
\mathbf{H}^{l+1} \ &=\ GTI\left(\mathbf{H}^{l} ,\mathbf{A}\right) = \mathbf{G}^{l+1} \ +\ \mathbf{T}^{l+1} \label{eq:3}
\end{align}
\noindent where $\mathbf{G}^{l+1}\in \mathbb{R}^{N\times d}$ is the generated graph representation, $\mathbf{T}^{l+1}\in \mathbb{R}^{N\times d}$ is the global attention feature matrix and $d$ is the dimension of the feature embedding.

(1) \textit{GCN:} The message passing functions of the general $GCN$ operator, acting on the local neighborhood of node $u$ at $l$-th layer, can be represented as follows: 
\begin{align}
\mathbf{m}_{u}^{l} \ \ &=\ AGG\left(\left\{\mathbf{m}_{uv}^{l} = \rho \left(\mathbf{h}_{u}^{l},\mathbf{h}_{v}^{l} ,\mathbf{h}_{e_{uv}}^{l}\right) ,\ v\in \mathcal{N}( u)\right\}\right) \label{eq:5}\\
\mathbf{g}_{u}^{l+1}  &=\ \phi\left(\mathbf{h}_{u}^{l} ,\ \mathbf{m}_{u}^{l}\right) \label{eq:6}
\end{align}
\noindent where $\rho$, $AGG$, and $\phi$ are differentiable functions. The message construction function $\rho$ constructs a message for node $u$ by integrating the node $u$ feature $\mathbf{h}_{u}^{l}$, features of its neighbors $\mathbf{h}_{v}^{l}$, and the edge features $\mathbf{h}_{e_{uv}}^{l}$. The $AGG$ is a permutation invariant function that aggregates all messages directed towards node $u$. 
In essence, the $AGG$ function executes MIL-manner operations within a graph's local neighborhood. 
The resulting feature $\mathbf{g}_{u}^{l+1}$ of node $u$ is then updated by merging the original node feature $\mathbf{h}_{u}^{l}$ and the aggregated message $\mathbf{m}_{u}^{l}$ via the update function $\phi$.
As the choice of these GCN related functions is flexible, we adopt the generalized graph convolution $GENConv$ from the DeeperGCN~\cite{li2020deepergcn}. The corresponding message passing functions are defined as follows:
\begin{align}
\mathbf{m}_{uv}^{l} &=\ ReLU\left(\mathbf{h}_{v}^{l} +\mathbf{1}\left(\mathbf{h}_{e_{uv}}^{l}\right) \cdot \mathbf{\ h}_{e_{uv}}^{l}\right) \ +\ \epsilon \label{eq:7} \\
\mathbf{m}_{u}^{l} \ \ &=\ \sum\limits_{v\in \mathcal{N}(u)} \frac{exp\left( \beta \mathbf{m}_{uv}^{l}\right)}{\sum\limits_{v\in \mathcal{N}(u)} \ exp\left( \beta \mathbf{m}_{uv}^{l}\right)} \cdot \mathbf{m}_{uv}^{l} \label{eq:8}\\
\mathbf{g}_{u}^{l+1}  &=\ \phi\left(\mathbf{h}_{u}^{l} ,\ \mathbf{m}_{u}^{l}\right)
=\ MLP\left(\mathbf{h}_{u}^{l} \ +\mathbf{m}_{u}^{l}\right)\label{eq:9}
\end{align}
The message is constructed by a ReLU activation function with the neighboring node feature $\mathbf{h}_{v}^{l}$ and the associated edge feature between node $u$ and $v$ where $\mathbf{1}(\cdot)$ is an indicator function. A small positive constant ($\epsilon=10^{-7}$) is added to the ReLU activation function output to ensure positive feature values for the numerical stability. The resulting messages from neighboring nodes are summed with weights by the SoftMax function where hyper-parameter $\beta$ denotes the inverse temperature. This aggregation method concentrates on the local instance interactions. Finally, the update function is structured as a two-layer MLP. These configurations ensure an effective feature transformation and message propagation.

(2) \textit{Global Attention:} While GNNs can be used to describe the entire WSI graph, they can be  constrained for long-range dependency characterization due to the limited receptive field.
Although an increase in a GNN depth could be a potential remedy, it can result in indistinguishable node representations, an issue known as over-smoothing or over-squashing. To alleviate these problems, we implement a global attention modules in parallel to the GCN (Fig.~\ref{fig:method}). This design enhances the ability to identify discriminating node representations from the entire WSI graph. 
Specifically, the global attention layer employs the self-attention mechanism, with its formulation given below:
\begin{align}
GlobalAttn(\mathbf{Q} ,\mathbf{K} ,\mathbf{V}) \ =\ softmax\left(\frac{\mathbf{QK}{^{T}}{}}{\sqrt{d_{q}}}\right)\mathbf{V}
\end{align}
where feature representations $\mathbf{Q} ,\mathbf{K} ,\mathbf{V}$ are calculated by projecting instance feature matrix $\mathbf{H}$ using distinct three weight matrix $\mathbf{W}_{i} \ \in \ \mathbb{R}^{d\times d_{i}}$. While the self-attention mechanism in the original transformer is effective and well-suited for this scenario, its $O(N^2)$ computational complexity limits its ability to process a large number of input instances efficiently. To address this limitation, we leverage FlashAttention (FA)~\cite{flashattention} to fully harness the potential of the multi-head self-attention mechanism without information loss or an expensive computational cost.
Integrating global feature embeddings with those from the GCN branch, we produce effective and expressive WSI representations that are able to capture the global contextual information and the local neighbor interactions.

After the feature processing via GTI blocks, a straightforward attention-based MIL pooling~\cite{attentionpooling} strategy is used for feature aggregation in the downstream phase in Fig.~\ref{fig:method}. The resulting bag-level representation $\mathbf{h_{bag} \ } \in \ \mathbb{R}^{1\times d}$ is computed by the weighted average of the instance representations by the attention scores $\alpha$ as follows:
\begin{align}
\mathbf{h_{bag}} \ =\ \mathbf{\alpha }^{T}\mathbf{H}^{L}
\label{eq:11}
\end{align}
In the final phase, the bag-level feature $\mathbf{h_{bag}}$ is provided to the MLP layer to achieve the final bag-level classification.

%% file: chapter/experiment.tex
\section{Experiments}

\subsection{Datasets}
To demonstrate the efficacy of our novel IGT framework, we conduct experiments and compare our method with SOTA methods on three widely used public datasets: TCGA-NSCLC (The Caner Genome Atlas Non-Small Cell Lung Cancer), TCGA-RCC (Renal Cell Carcinoma) and BRIGHT~\cite{brancati2022bracs}. We use the official data split if it is available, otherwise, we split the train, validation, and test sets by an  ratio of 6.5:1.5:2.0. 
All WSIs in these datasets are cropped at $20\times$ magnification.

\noindent\textbf{TCGA-NSCLC} is a lung cancer dataset and includes two distinct cancer subtypes: Lung Adenocarcinoma (LUAD) and Lung Squamous Cell Carcinoma (LUSC). It has 1,043 diagnostic digital WSIs with 531 and 512 WSIs of LUAD and LUSC, respectively. We follow the same random split for DSMIL study~\cite{dsmil}.

\noindent\textbf{TCGA-RCC} is a kidney cancer dataset and consists of 940 WSIs. Specifically, there are 121 WSIs of 109 Kidney Chromophobe Renal Cell Carcinoma (TCGA-KICH) cases, 519 WSIs of 513 Kidney Renal Clear Cell Carcinoma (TCGA-KIRC) cases, and 300 WSIs of 276 Kidney Renal Papillary Cell Carcinoma (TCGA-KIRP) cases. 

\noindent\textbf{BRIGHT} is a breast cancer dataset and contains 503 diagnostic slides across six breast tumor subtypes: Pathological Benign (PB), Usual Ductal Hyperplasia (UDH), Flat Epithelia Atypia (FEA), Atypical Ductal Hyperplasia (ADH), Ductal Carcinoma in Situ (DCIS), and Invasive Carcinoma (IC). 
We use the official data split, where 423 WSIs are for training and 80 WSIs for testing.

\subsection{Implementation Details}
In the graph construction phase, background patches with a saturation level of less than 15 are discarded. The processed 1024-dimensional feature vector $h_{i} \in \mathbb{R}^{1024}$ is downscaled to 256 and assembled for the node feature matrix $H\in R^{N\times 256}$~\cite{clam,transmil}, before being taken as input. For model training, the cross-entropy loss function is utilized, and the batch size is set to 1. We adopt the Rectified Adam optimizer~\cite{radam} for optimization with a weight decay of 1e-5. We train the IGT framework for 40 epochs on both TCGA-NSCLC and TCGA-RCC datasets, and for 30 epochs on BRIGHT dataset. The learning rate starts at 1e-3, decaying to 1e-4 at epoch 20 for TCGA-NSCLC, and at epoch 15 for TCGA-RCC and BRIGHT. We employ two GTI blocks for TCGA-NSCLC and TCGA-RCC, and three GTI blocks for BRIGHT. 
All models are implemented by PyTorch 2.0, and executed on an NVIDIA GeForce RTX 3090Ti GPU.

\begin{table*}[t]
\caption{Comparison of accuracy and AUROC on three public datasets. The reported metrics are presented as percentages and averaged for three times.
Our IGT framework consistently outperforms existing state-of-art MIL methods.}
\centering
\begin{tabular}{ l c c c c c c c}
\toprule
\multirow{2}{*}{Method} &\multicolumn{2}{c}{TCGA-NSCLC} &\multicolumn{2}{c}{TCGA-RCC}  &\multicolumn{2}{c}{BRIGHT-6class}\\
\cmidrule(lr){2-3} \cmidrule(lr){4-5} \cmidrule(lr){6-7}
    &ACC(\%) &AUC(\%)  &ACC(\%) &AUC(\%) &ACC(\%) &AUC(\%)\\
    \hline
Mean-pooling &77.6 &86.2  &82.3  &94.2  &26.1 &64.1\\
Max-pooling &79.0 &85.8  & 84.0 &96.1  &29.3 &66.0\\
ABMIL\cite{attentionpooling} &84.1 &91.3 &86.8 &97.1  &30.8  &67.0 \\
DSMIL\cite{dsmil} &86.0 &93.9     &87.7 &97.7  &36.4 &72.5 \\
CLAM-SB\cite{clam} &85.5 &90.9     &88.5 &98.0   & 33.1 &69.1  \\
CLAM-MB\cite{clam} &87.9 &92.9     &89.9  &97.9   &38.1 &71.7  \\
TransMIL\cite{transmil} &89.3 &94.2    &90.2 &97.7 &39.6 &71.8 \\
\hline
GCN-ABMIL\cite{liang2023interpretable} &87.3 &94.4  &89.2 &97.6  
&33.4 &68.1 \\
Patch-GCN\cite{patchgcn} &88.8 &95.0  &89.7 &98.1  &38.2 &71.2\\
GTP\cite{GTP} &90.5 &95.8  &91.4 &97.7 &40.8 &72.9 \\ 
\hline
IGT (Ours) &\textbf{91.6} &\textbf{96.7}    &\textbf{92.4} &\textbf{98.4}   &\textbf{43.4} &\textbf{74.5} \\
\bottomrule
\end{tabular}

\label{tab:1}
\end{table*} 

\subsection{Results}
\subsubsection{Performance comparison with the SOTA methods:}
We compare the proposed IGT with ten baselines: 
Seven of them are none graph based methods, including max/mean-pooing, ABMIL~\cite{attentionpooling}, DSMIL~\cite{dsmil}, CLAM-SB~\cite{clam}, CLAM-MB~\cite{clam} and TransMIL~\cite{transmil}.
Three of them are graph-based MIL methods, including GCN-ABMIL~\cite{liang2023interpretable}, PatchGCN~\cite{patchgcn} and GTP~\cite{GTP}.
Note both TransMIL and GTP use Transformers. 
We chose overall accuracy (ACC) and area under receiver operating characteristic curve (AUROC) as the evaluation metrics.


As illustrated in Table~\ref{tab:1}, our IGT framework surpasses current SOTA methods. To be specific, compared with the best performing graph-based method, GTP, our method achieves a 1.1\% improvement in accuracy and a 0.9\% increase in AUROC for the binary classification on the TCGA-NSCLC dataset. In multi-class classification, our method shows a 1.0\% improvement in accuracy and a 0.7\% increase in AUROC for the TCGA-RCC dataset, and a 2.6\% improvement in accuracy with a 2.5\% increase in AUROC on the BRIGHT dataset. Similarly, when compared with the leading non-graph-based method, TransMIL, our method shows a substantial 2.2\%-3.8\% improvement in accuracy and a 0.7\%-2.7\% enhancement in AUROC.
In conclusion, our graph-transformer-based method significantly outperforms current both graph-based and transformer-based approaches, indicating the advantages of integrating local neighborhood information with global context for enhanced performance. 

\subsubsection{Ablation Studies}

To demonstrate the efficacy of the developed GTI block and investigate the necessity of model components, we conduct a ablation study to quantify the separate benefit of the individual global-attention and GCN module using ABMIL and DSMIL as the aggregation modules. 
As shown in Table~\ref{tab:2}, compared with GTI block without self-attention, our GTI achieves a 2.4\% to 5.7\% improvement in accuracy. It proves that the self-attention mechanism in our GTI captures pairwise correlation across all instances and thus improves the performance. 
In comparing our GTI block with GTI block without the GCN branch, our GTI block achieves a 3.7\% to 5.6\% increase in accuracy. It shows the necessity of spatial information for WSI analysis. 

An interesting finding is that the method equipped exclusively with the global attention module exhibit inferior performance compared to those only utilizing  the GCN. This discrepancy can be attributed to the lack of spatial information when directly applying the self-attention mechanism for WSI analysis. 

\begin{table*}[h]
\caption{An ablation study conducted to evaluate the importance of each component within the GTI block, utilizing ABMIL and DSMIL as the base aggregation models.}
\centering
\begin{tabular}{ c c c c c c c c c c}
\toprule
\multirow{2}{*}{Aggregator} &\multirow{2}{*}{Backbone} &\multicolumn{2}{c}{TCGA-NSCLC} &\multicolumn{2}{c}{TCGA-RCC}  &\multicolumn{2}{c}{BRIGHT-6class}\\
    \cmidrule(lr){3-4} \cmidrule(lr){5-6} \cmidrule(lr){7-8}
& &ACC(\%) &AUC(\%)  &ACC(\%) &AUC(\%) &ACC(\%) &AUC(\%)\\
    \hline
\multirow{2}{*}{ABMIL} &   -       &84.1 &91.3 &86.8 &96.1 &30.8 &67.0 \\
                       & GTI w/o Attn  &89.2 &95.2 &88.8 &98.1 &38.7 &72.5\\
                       & GTI w/o GCN  &86.0 &93.1 &87.8 &98.0 &38.1 &71.2\\      
                       & GTI      &\textbf{91.6} &\textbf{96.7} &\textbf{92.4} &\textbf{98.4} &\textbf{43.4} &\textbf{74.5}\\
\hline
\multirow{2}{*}{DSMIL} &   -       &86.0 &93.9 &87.7 &97.7 &36.4 &72.5\\
                       &GTI w/o Attn  &87.9 &94.3 &89.8 &98.1 &39.0 &73.7\\
                       &GTI  w/o GCN  &87.4 &95.2 &88.7 &98.0 &37.4 &\textbf{73.8}\\
                       & GTI      &\textbf{91.1} &\textbf{95.5} &\textbf{91.7} &\textbf{98.5} &\textbf{42.9} &73.4 \\
\bottomrule
\end{tabular}
\label{tab:2}
\end{table*}

%% file: chapter/conculsion.tex
\section{Conclusion}
In this paper, we introduce a new integrative graph-transformer framework, IGT, that simultaneously captures the context-aware relational features from local tissue regions and global WSI representations across instance embeddings for histopathology WSI classification. We integrate the graph convolutional network with a global attention module to construct the Graph-Transformer Integration block. Specifically, the graph convolutional network explores the local neighbor interactions and the multi-head self-attention model captures the long-range dependencies from all instances. The efficacy of the developed framework is manifested with three public WSI datasets. When compared with multiple state-of-the-art methods, our method consistently presents a superior performance, suggesting its promising potential to support computational histopathology analyses.
